\documentclass[preprint,showkeys,aps,prb]{revtex4}
\usepackage{amsmath}
\usepackage[dvips]{graphicx}
\usepackage{setspace}
\begin{document}
\title{
Canonical Temperature Control by Molecular Dynamics\cite{b1,b2}
}

\author{
William Graham Hoover and Carol Griswold Hoover       \\
Ruby Valley Research Institute  \\
HC 60 Box 601    \\
Ruby Valley, NV 89833                             \\
}

\date{\today}


\vspace{0.1cm}

\begin{abstract}
``Pedagogical derivations for Nos\'e's dynamics can be developed in two different ways, (i)
by starting with a temperature-dependent Hamiltonian in which the variable $s$ scales the
time or the mass, or (ii) by requiring that the equations of motion generate the canonical
distribution including a Gaussian distribution in the friction coefficient $\zeta$. Nos\'e's
papers follow the former approach. Because the latter approach is not only constructive and
simple, but also can be generalized to other forms of the equations of motion, we illustrate
it here. We begin by considering the probability density $f(q,p,\zeta)$ in an extended phase
 space which includes $\zeta$ as well as all pairs of phase variables $q$ and $p$. This density
$f(q,p,\zeta)$ satisfies the conservation of probability (Liouville's Continuity Equation)''
$$
(\partial f/\partial t) + \sum (\partial (\dot q f)/\partial q) + \sum (\partial (\dot p f)/\partial p)
+ \sum (\partial (\dot \zeta f)/\partial \zeta) = 0  \ .
$$
The multi-authored ``review''\cite{b1} motivated our quoting the history of Nos\'e and
Nos\'e-Hoover mechanics, aptly described on page 31 of Bill's 1986 {\it Molecular Dynamics} book,
reproduced above\cite{b2}.
\end{abstract}

\maketitle

\section{Introduction}

In 1984 Shuichi Nos\'e discovered a canonical form of molecular dynamics\cite{b3,b4} consistent
with Gibbs' canonical ensemble probability density, $ f \propto e^{-{\cal H}(q,p)/kT}$. Bill was
struck by the revolutionary nature of Nos\'e's papers. As a result he arranged to attend a workshop meeting at
Orsay, just outside Paris, where he and Nos\'e were scheduled to talk about molecular dynamics.
A stroke of luck brought Bill and Shuichi together a few days earlier, purely by accident, at a Paris
train station. Bill identified Shuichi by his large suitcase bearing the label ``NOSE''. The two
arranged to meet for technical discussions on a bench in front of the Notre Dame Cathedral. Bill
brought with him a list of about a dozen questions for Shuichi. Both came away with a better
understanding of Nos\'e's discovery. Shuichi's two papers were difficult reading for Bill.  They
involved ``scaling the time'' so as to provide the Gaussian canonical distribution of velocities
along with the Boltzmann-factor $\propto e^{-\Phi/kT}$ probability density for the coordinates and
$e^{-K/kT}$ for the scaled momenta, $\{ \ (p/s) \ \}$. All
this Nos\'e  accomplished by introducing a time-scaling variable $s$ along with its conjugate momentum
$p_s$. Nos\'e's two papers, with about 20 pages of algebra, provided a novel and highly-productive
 connection of molecular dynamics to Gibbs' canonical statistical mechanics.

The concept of time-scaling, relating ``real'' time to ``virtual'' time, made reading Nos\'e's
papers a heavy lift. To simplify this task Bill hit upon the idea of applying Nos\'e's ideas to
a simple example problem, the one-dimensional harmonic oscillator. He began a manuscript\cite{b5}
in Orsay and completed it in Lausanne after the Orsay workshop, thanks to a kind invitation from
Philippe Choquard to visit his home and the Lausanne laboratory. Along with Harald Posch and Franz
Vesely, Bill pursued the oscillator problem further in Vienna\cite{b6}. They found periodic, toroidal,
and chaotic multifractal solutions of the oscillator equations. The simplest case considered is
described by three ordinary differential equations (enough for chaos) giving the evolution of the
coordinate $q$, momentum $p$, and friction coefficient $\zeta$ :
$$
\{ \ \dot q = p \ ; \ \dot p = -q - \zeta p \ ; \ \dot \zeta = p^2 - 1 \ \} \ 
[ \ {\rm Nos\acute{e}'-Hoover \ Oscillator} \ ] \ .
$$
It is easy to use the continuity equation to confirm that the steady-state canonical distribution
$f \propto e^{-(q^2+p^2+\zeta^2)/2}$ is consistent with these Nos\'e-Hoover motion equations. 

In the years since 1984 the Nos\'e-Hoover motion equations have become the standard algorithmic technique
for isothermal simulations. Tens of thousands of citations of Nos\'e and Hoover's papers testify to
their value in stimulating additional thermostat research, both at and away from, equilibrium.  There are
occasional setbacks. See particularly
the relatively recent Reference 1 responsible for the present work. Although described as a review that
article entirely misstates the history of thermostatted mechanics and ignores the vast computational
literature on applications to chaotic irreversible processes. In addition to our own work\cite{b7} see
also the fundamental contributions of Dettmann, Evans, and Morriss \cite{b8,b9,b10}. Among many other
developments  Dettmann and Morriss discovered a Hamiltonian ${\cal H}_{\rm DM}=s{\cal H}_{\rm Nos\acute{e}}$
which generates the Nos\'e-Hoover equations directly, without the need for a separate time-scaling step.
Demonstrating this connection is an interesting exercise for the reader.

It is particularly noteworthy that nonequilibrium simulations are the primary beneficiary of all the
work on deterministic thermostats. Isoenergetic, isokinetic, and isobaric thermostats all have provided
new algorithms linking time-reversible equations of motion to irreversible simulations. The papers by
Bauer, Bulgac, and Kusnezov provide a useful guide to the construction of new algorithms\cite{b11}.

\section{An Interesting Toy Problem Example}

\begin{figure}
\includegraphics[width=3in,angle=+90.]{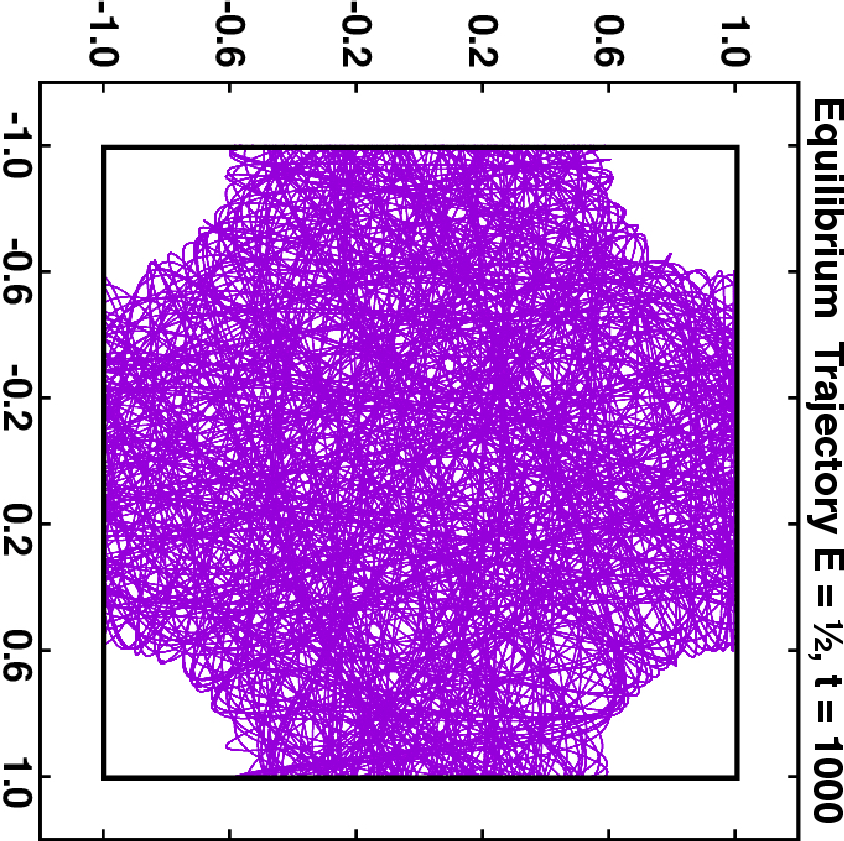}
\caption{A million-timestep trajectory with $dt = 0.001$ and periodic boundaries in the $x$ and $y$
directions. The initial condition is $p = (0.6,0.8)$ with $(x,y) = (0,0)$. The total energy agrees
with the initial to eleven-figure accuracy at the conclusion of the run. The maximum potential
energy of $1/2$ occurs along four quarter-circles centered at $(\pm 1, \pm1)$, with radii $0.398878 =                 
\sqrt{1-2^{-1/4}}$.
}
\end{figure}

Here we provide an interesting toy problem\cite{b7} suited to illustrating both approaches to canonical simulations,
scaling the time, and introducing time-reversible friction. The two subsections following these different
approachs can provide {\it identical} $(x,y)$ trajectories, but with {\it different} $(p_x,p_y)$ momenta. The
system explored here is a one-body ``wanderer'' problem remeniscent of the Einstein cell model of solid state physics.
The two-dimensional $(x,y)$ motion takes place within a periodic square of sidelength 2, centered on the
origin $(x,y)= (0,0)$. Four fixed scatterers, at the corners $(x,y) = (\pm 1,\pm 1)$, influence the motion of the wanderer
particle. The potential furnished by the four scatterers has the very smooth form $\phi(r<1) = (1-r^2)^4$. The
conventional Hamiltonian equations of motion are
$$
\{ \ \dot x = p_x \ ; \ \dot y = p_y \ ; \ \dot p_x = \sum 8dx(1-r^2)^3 \ ; \ \dot p_y =\sum 8dy(1-r^2)^3 \ \} .
$$ 
The sums include only those scatterers, if any, with deviations $r = \sqrt{(dx^2 + dy^2)}$ from the wanderer
less than unity. For simplicity we take an initial condition with a (conserved) energy of 0.5 :
$(x,y,p_x,p_y) = (0,0,0.6,0.8)$. Let us summarize the two approaches to the Nos\'e-Hoover equations, Nos\'e's,
based on his 1984 papers\cite{b3,b4}, and Hoover's, based on his 1985 work\cite{b5}.

\subsection{Nos\'e's Approach: Scaling the Time}

The first step in Nos\'e's derivation is to augment the conventional Hamiltonian $K + \Phi$, with $(s,p_s)$,
the time-scaling variable $s$ and its conjugate momentum $p_s$:
$$
{\cal H} = (K/s^2) + \Phi + (p_s^2/2) + \ln(s) \ [ \ {\rm Nos\acute{e}'s \ Hamiltonian} \ ] \ .
$$
Next, the resulting equations of motion, $(\dot x,\dot y,\dot s, \dot p_x,\dot p_y,\dot p_s)$: 
$$
\{ \ \dot x = (p_x/s^2) \ ; \ \dot y = (p_y/s^2) \ ; \ \dot s = p_s \ \} \ {\rm Coordinates} \ ;
$$
$$
\{ \ \dot p_x = F_x \ ; \ \dot p_y = F_y \ ; \ \dot p_s = (p_x^2 + p_y^2)/s^3 - (1/s) \ \} \ {\rm Momenta} \ ,
$$

are multiplied by $s$, ``scaling the time''. Third, and last, the ``scaled momenta'', $(p_x/s)$ and $(p_y/s)$,
are replaced by $p_x$ and $p_y$. The resulting equations of motion are the Nos\'e-Hoover equations:
$$
\{ \ \dot x = p_x \ ; \ \dot y = p_y \ ; \ \dot p_x = F_x - \zeta p_x \ ;
\ \dot p_y = F_y - \zeta p_y \ ; \ \dot \zeta = K - 1/2 \ \} \ [ \ {\rm Nos\acute{e}-Hoover}\ ] 
$$
Despite the smooth nature of the potential function, solutions of the Nos\'e equations are typically stiff.
{\bf Figure 2} illustrates the evolution of the time-scaling factor $s$ for the cell-model problem of
{\bf Figure 1}.

\begin{figure}
\includegraphics[width=3in,angle=-90.]{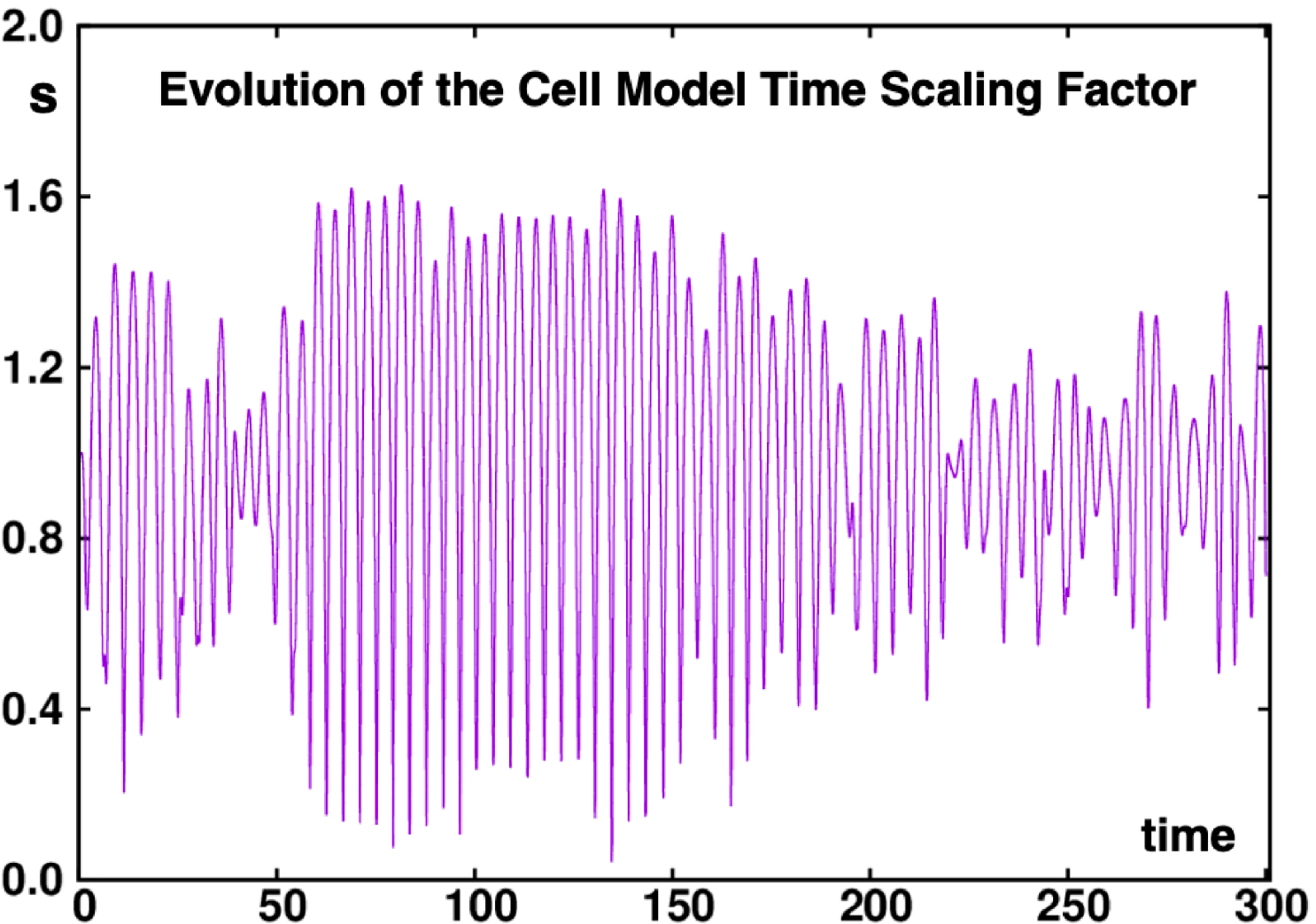}
\caption{
Evolution of the time-scaling factor $s$ for the Toy Model cell model for the initial 300 000 fourth-order
Runge-Kutta timesteps of 0.001.
}
\end{figure}

Nos\'e's three-step ``derivation'' of the Nos\'e-Hoover equations looks like magic rather than straightforward
mechanics. His highly original search for a time scale linking isoenergetic and isothermal motion equations
used three unconventional steps in scaling the time. In their February 2006 Physics Today obituary of Shuichi,
Yosuke Kataoka and Michael L. Klein recall that his two 1984 articles were ``somewhat delayed by referees
who had difficulty accepting the new and highly original formulation''.

By contrast, Hoover's derivation of the Nos\'e-Hoover equations relies on the phase-space continuity equation,
an analog of Liouville's Theorem, a standby of conventional statistical mechanics. We summarize that next.

\subsection{Hoover's Approach: The Continuity Equation}

After a couple of weeks of study, in France and Switzerland, Hoover found a straightforward path to both the
isothermal and the isobaric Nos\'e-Hoover equations. The basis is the continuity equation for the conservation
of probability in phase space, Liouville's Theorem. For simplicity we illustrate the isothermal steps for a
single degree of freedom. We begin with the assumption that the motion equations, $\{ \  \dot p = F - \zeta p \ \}$,
include a friction coefficient depending on the phase variables, $\zeta(q,p)$.
We also assume that an exponential form, $e^{-{\cal F}(\zeta)}$, multiplies the conventional canonical Gibbs' distribution
$f(q,p,t) \propto e^{-{\cal H}/kT}$. Suppose that the equations of motion need nothing more than a
linear friction coefficient, $\dot p = F -\zeta p$, to acquire an extended canonical solution,
$f \propto \exp[-\Phi/T - K/T - {\cal F}(\zeta)]$. For a steady-state Liouville's Theorem, the continuity equation in the extended $(q,p,\zeta)$ phase space, implies that
$(\partial f/\partial t)$ vanishes:
$$                                                                                                                                          (\partial(\dot qf)/\partial q) + (\partial(\dot pf)/\partial p) + (\partial(\dot \zeta f)/\partial \zeta) = 
-(\partial f/\partial t) \equiv 0 \ .
$$
Two relations describing the flow in $(q,p)$ space provide the Nos\'e-Hoover distribution function. For simplicity
we write the relations for a single canonical pair and choose the temperature, Boltzmann's constant, mass, and the
relaxation time of the frictional force, $-\zeta p$ all  equal to unity:
$$
 (\partial(\dot qf)/\partial q) + (\partial(\dot pf)/\partial p) =
-(\partial(\dot \zeta f)/\partial \zeta) = -\dot \zeta (\partial f/\partial \zeta) = -(d{\cal F}/d\zeta )\dot \zeta f \ ;
$$
$$
(\partial(\dot qf)/\partial q) + (\partial(\dot pf)/\partial p) =
pFf + (F - \zeta p)(-pf) - \zeta f = \zeta(p^2 - 1)f \ .
$$

The joint solution of these two flow relations, ${\cal F}(\zeta)=(\zeta^2/2)$ and $\dot \zeta = p^2 - 1$, gives Gibbs' canonical
distribution, augmented by a Gaussian distribution of the friction coefficient :
$$
f\propto \exp[ -{\cal H} - (\zeta^2/2) \ ] \longrightarrow \dot \zeta = (p^2 - 1) \ ,
$$
$$
\dot p = F - \zeta p \ ; \ \dot \zeta \propto K - 1/2 \ \} \ [ \ {\rm Nos\acute{e}-Hoover \ Equations} \ ] \ .
$$
This is the simplest form of the Nos\'e-Hoover algorithm and its one-step derivation is arguably the
simpler of the two routes to this time-reversible deterministic canonical dynamics.

From the standpoint of simplicity Hoover's assumption of a friction coefficient (which turns out to
be the momentum $p_s$ conjugate to Nos\'e's $s$) is preferable to the time-scaling Hamiltonian and
the redefinition of momentum in Nos\'e's work. It is noteworthy too, that a dozen years later, Dettmann
and Morriss found a Hamiltonian which automatically accomplishes Nos\'e's program\cite{b8,b9}.

\section{Summary}

\begin{figure}
\includegraphics[width=3in,angle=-90.]{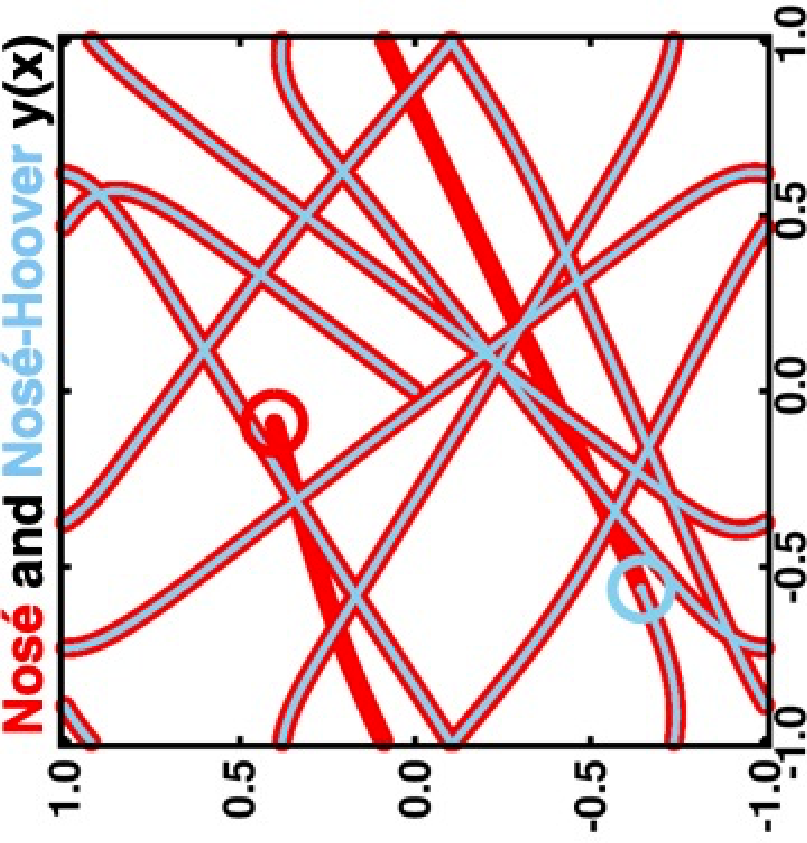}
\caption{
Evolution of the cell model coordinates $(x,y)$ from Nos\'e's Hamiltonian (red, and concluding
at the open red circle just above the origin) are compared to those from the Nos\'e-Hoover motion
eequations (blue, and ceasing at the blue open circle at lower left). The two approaches follow
{\it identical trajectories}, but at {\it different rates}. Here the results of 20 0000 fourth-order
Runge-Kutta timesteps are displayed for both sets of motion equations with initial conditions
$(x,y,s,p_x,p_y,p_s=\zeta) = (0,0,1,0.6,0.8,0)$.
}
\end{figure}

We have outlined two approaches to the Nos\'e-Hoover motion equations. Both were well established
in 1984. Both have stimulated the development of deterministic thermostats, with nonequilibrium
steady-state simulations generating fractal phase-space distributions. The 1984 and 1986 Paris workshops
stimulated a simple example problem\cite{b12}. By 1987 deterministic time-reversible thermostatting
was used to resolve Loschmidt's paradox for thermostatted steady states\cite{b13}. {\bf Figure 3} shows
two solutions of the motion equations with initial conditions $(x,y,s,p_x,p_y,p_s) = (0,0,1,0.6,0.8,0)$.
Comparing the two shows that the Nos\'e version is ``stiffer'' than the Nos\'e-Hoover. The culprit is
the small denominator in the differential equation for $p_s: \dot p_s = (p^2/s^3) - (1/s)$. See the
discussion in pages 123-126 of our book of Kharagpur lectures\cite{b7}.

This toy model problem presents the opportunity for future work studying heat transfer between the
horizontal and vertical degrees of freedom and the challenge of displaying graphic evidence for strange
attractors in a six-dimensional phase space.

\section{Acknowledgment}

We are grateful to Kris Wojciechowski for alerting us to Reference 1 and for his support and
encouragement, leading to this comment. We also thank Hesam Arabzadeh at the University of
Missouri for several stimulating emails.

\section{Appendix}
Two aspects of programming thermostatted mechanics for the cell model are worth describing here.
Looping over the four fixed scatterers in computing the forces or the energy is simplest with stored
arrays of the scatterers' $x$ and $y$ coordinates; \\
{\tt
      dimension xj(4),yj(4)    \\
      xj(1) = +1 ; yj(1) = +1 ; xj(2) = -1 ; yj(2) = +1  \\
      xj(3) = -1 ; yj(3) = -1 ; xj(4) = +1 ; yj(4) = -1  \\
}
To illustrate the use of these arrays consider the computation of the energy: \\
{\tt
phi = 0 ; do j = 1,4                       \\
dx = x - xj(j) ; dy = y - yj(j) ; rr = dx*dx + dy*dy \\
if(rr.lt.1) phi = phi + (1 - rr)**4 ; end do   \\
}
After each Runge-Kutta integration step the four checks of the periodic boundaries
need to be implemented: \\
{\tt
if(x.gt.+1) x = x - 2  \\
if(x.lt.-1) x = x + 2  \\
if(y.gt.+1) y = y - 2  \\
if(y.lt.-1) y = y + 2  \\
}

\pagebreak

\end{document}